# Exploiting peer group concept for adaptive and highly available services

Muhammad Asif Jan
*Centre for European Nuclear Research (CERN) Switzerland*

Fahd Ali Zahid, Mohammad Moazam Fraz
*Foundation University, Islamabad, Pakistan*

Arshad Ali
*National University of Sciences and Technology, Rawalpindi, Pakistan*

This paper presents a prototype for redundant, highly available and fault tolerant peer to peer framework for data management. Peer to peer computing is gaining importance due to its flexible organization, lack of central authority, distribution of functionality to participating nodes and ability to utilize unused computational resources. Emergence of GRID computing has provided much needed infrastructure and administrative domain for peer to peer computing. The components of this framework exploit peer group concept to scope service and information search, arrange services and information in a coherent manner, provide selective redundancy and ensure availability in face of failure and high load conditions. A prototype system has been implemented using JXTA peer to peer technology and XML is used for service description and interfaces, allowing peers to communicate with services implemented in various platforms including web services and JINI services. It utilizes code mobility to achieve role interchange among services and ensure dynamic group membership. Security is ensured by using Public Key Infrastructure (PKI) to implement group level security policies for membership and service access.

## 1. INTRODUCTION

Information and message flow in distributed systems is not guaranteed where as entities in distributed systems, clients and application, always desire some level of quality of service assurance. A realistic distributed system shall be able to adapt to faulty conditions at runtime and be able to maintain a threshold in the performance and the availability. One of the possible approaches to handle inherently unreliable nature of distributed system is to provide selectively redundant self organising peer services. Peer services shall be able to adapt to various high load and failures states and be able to provide assurance of service availability. Thus adaptive redundancy can result in a greater availability and performance of distributed services.

Services can be seen as roles which peers can play in a peer group and are accessible in context of a peer group. Services are not bound to specific endpoints, thus allowing them to recover from specific peer failures and also to start new services in high load conditions. A strategy for role interchange can be configured to allow peers to make decisions based on specific conditions and assume additional responsibilities to ensure that the group services are maintained to a desired level of availability. All peers in the systems can be identical and there is no additional requirement for roles that they have to play in the peer group other then imposed by local network policies.

We propose a peer services based framework also known as CONSCIENTIA[1]. It is based on the JXTA[2] open P2P communication protocol stack. Project JXTA was started at Sun Microsystems in 2001. JXTA defines a set of protocols that can be implemented by peers to communicate and collaborate with other peers implementing the JXTA protocol stack. It tries to standardize messaging systems, specifically in peer-to-peer systems, by defining protocols, rather than implementations. A strict adherence to XML messaging format allows services and peers to intercommunicate. Each peer service is represented by a ID that uniquely identifies peer or service on the network.

## 2. CONSCIENTIA FRAMEWORK ARCHITECTURE

CONSCIENTIA follows the Service Oriented Architecture [3]. Services register themselves to a super group where they are discovered by clients.

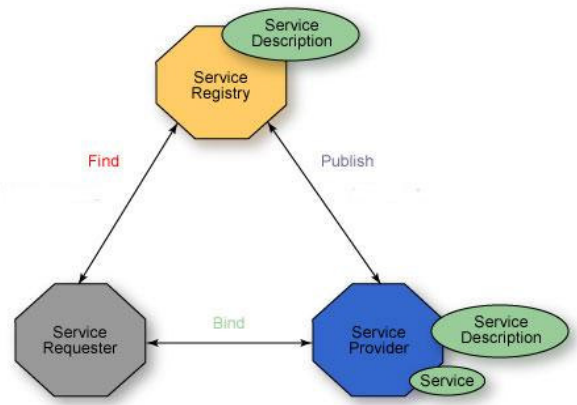

Figure (1) SOA architecture Diagram

Services are organized in peer groups. Groups are defined as per purpose and functionality of the group, for example a database service will be part of Database Group and a file access service can be accessed in File Group.

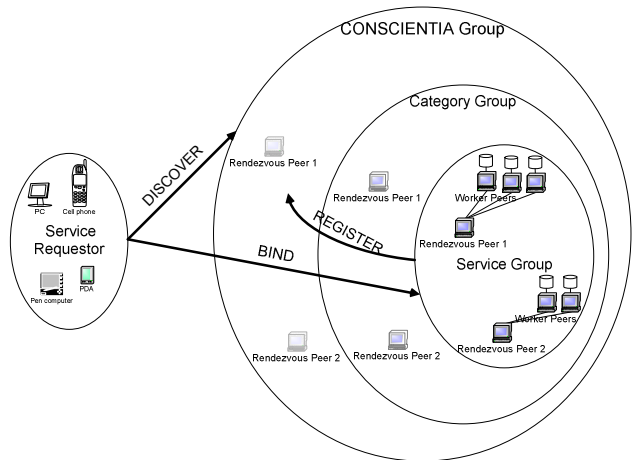

Figure(2) CONSCIENTIA SOA Diagram.

CONSCIENTIA group forms the super (default) group in this peer to peer framework. All other subgroups are part of CONSCIENTIA group; all the services are categorized and maintained in Category subgroups. Service categorization helps in scoping the search context and improves performance. This also allows us to move all functionality to specific peer groups and super-group supports common operations such as search and discovery. Every service is maintained in a Service Group native to that service.

## 3. CONSCIENTIA PEERS

The framework distinguishes three different types of peers.

- Rendezvous Peers.
- Worker Peers.
- Client peers.

### 3.1. Rendezvous (RV) Peers

The Rendezvous Peers (RV) are the super nodes of the CONSCIENTIA framework. These peers are I/O intensive peers and are selected on the basis of their network connectivity. All service providing peers (Worker Peers) must register (Join the group) themselves with a RV peer. Every peer group will have at least one RV peer. A peer group can have more than one RV peer to handle high load conditions. RV peers are responsible for routing messages to their registered worker peers. It also provides following additional services:

- Entry point Service
- Monitoring Service

Each RV will have a threshold determining the number of peers that can register with it. When this threshold is crossed then another RV peer is started. The new RV makes a subgroup and worker peers register with it. This helps in distributing client requests to multiple RV peers to improve performance. Since RV peers do not coordinate client-service matchmaking, so it is possible to start RV peer service on any arbitrary peer with good connectivity. Similar policies can be exercised by closing down additional RV peers in low load conditions.

### 3.2. Worker Peers

Worker peers are process intensive peers. The worker peer runs the worker service. The main functionality of the worker peer is governed by the worker service. Worker peers are the basic work house of the service i.e. if a Data mining service is being offered, then the databases are physically hosted on worker peers.

Worker peers register themselves with a RV peer and provide references to their communication channels i.e. pipes[4]. Clients search for services, by submitting search queries to peer group through RV peers. The client embeds references to its communication path, i.e. pipe ID[4], in the search query. This allows worker peer to extract the client pipe ID and directly communicate with the client.

When a worker peer fails to discover the CONSCIENTIA Group, it creates a new group and serves as the RV peer. Subsequent peers will find this group and join it. The RV peer may leave the group at any instance. New peers will come forward to take over the responsibility of maintaining the CONSCIENTIA Group when the RV leaves.

### 3.3. Client Peers

The client peers are simple peers providing an interface to the client to discover, query and get results from services. The discovery is made by first joining the CONSCIENTIA Group. After joining, the client peer discovers all the sub groups within this group. A list of the sub groups is maintained by the RV peers in the CONSCIENTIA Group. The subgroups are basically the categorization of the registered services, after a category has been selected the client discovers the service groups of this category and chooses are service.

The query can only be sent when the client has joined the service group and formatted the query according to the format specified by the service. The query format is embedded in the service group's a

Advertisement[5] and the client can extract it while discovering the service group.

When the client peer is in the CONSCIENTIA Group it is connected to the RV peers of the CONSCIENTIA group, when the client joins a sub group then it is handed over to the RV peers in that sub group. Similarly when the client joins the service group it is connected to the RV peer of that group. This brings the client directly to the entry point of the service and a query can be sent efficiently to the service.

## 4. CONSCIENTIA GROUPS

In CONSCIENTIA Peers are organized in three types of groups.

- CONSCIENTIA Group.
- Category Group.
- Service Group.

The groups are encapsulated within each other. A Service Group is contained in Category Group, A Category Group is within the CONSCIENTIA Group, or may be encapsulated in another Category Group as a sub category.

A client peer has to join the CONSCIENTIA group to use a service. After joining the group a client peer can discover all the sub groups. Each sub group is a service category. The client peer will join the desired service category and discover all the services within it.

### 4.1. CONSCIENTIA Group

This group is the starting point for all peers. This group acts as a virtual service registry, and any service being provided in the framework can be discovered from here.

Isolated peers existing on private networks can make their own CONSCIENTIA groups and run local services. When such private networks connect to each other the isolated CONSCIENTIA groups merge. Merging is basically the ability of peers to discover each other. The merging is virtual and has impact on the physical topology. Peers from one group can now discover and use the services of the other group. When the connection disappears the groups are reduced to their isolated existence.

In figure(3) two isolated networks are shown that maintain their own CONSCIENTIA groups and their local services. When a peer common to both networks joins the group the local CONSCIENTIA groups virtually merge into a single group.

### 4.2. Category Group

Categorization of services helps in scoping the search context and reduces network traffic. Category group consists of service groups categorized on basis of services they offer i.e. Database group for database related services, Gaming group for Multiplayer Gaming services etc. There is no restriction to the level of nesting within a category group i.e. Gaming group for Multiplayer Gaming services may have Xbox and Sony subgroups.

All services must be encapsulated in a Category group. If a relevant category group does not exist, then the worker peers of a service create the group assume responsibility of maintaining the group.

Identical Category groups merge when their local CONSCIENTIA groups merge. All the subcategories recursively merge.

### 4.3. Service Group

The service group is unique for each service. Client peers need to join this group to access a particular service.

This group runs three local services that give the framework essential attributes like self healing, self managing and load balancing. These services exist in their respective groups within the service group.

These groups are:

- Worker Group.
- Entry Point & Monitor Group (EPM).

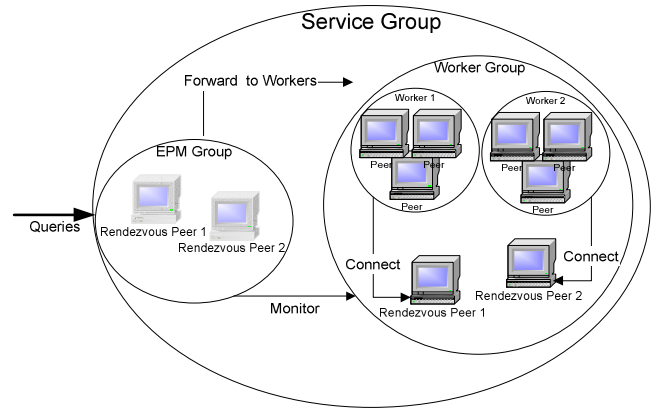

Figure (4) Service Group Architecture

#### 4.3.1. Worker Group

The worker group is composed of sub groups. Each sub group holds a cluster of worker peers. The concept of grouping the workers into sub groups helps in organizing the worker peers. The communication in a sub group is localized to that network space, this helps reduce the overall network traffic.

In Figure(4) a worker group is shown having two clusters of worker peers arranged in sub groups, "worker 1" and "worker 2". Both the groups have separate RV

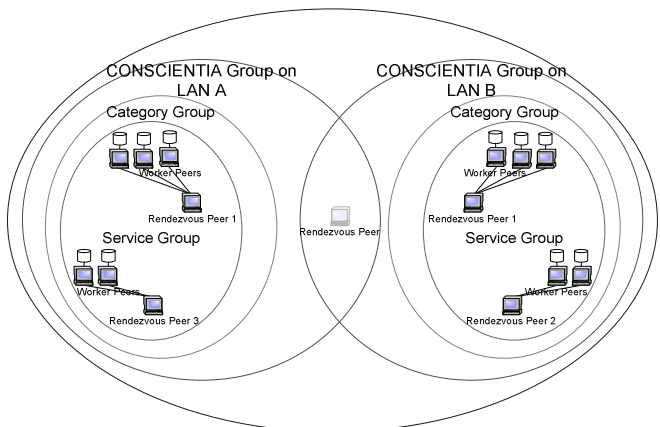

Figure (3) Merging of Conscientia Groups

peers "Rendezvous Peer 1" and "Rendezvous Peer 2" respectively.

### 4.3.2. Entry Point and Monitor Group (EPM)

The EMP group consists of only RV peers. As mentioned earlier that all the RV peers run monitoring service and entry point service. Theses services incorporate a heartbeat mechanism and query routing which involves all the RV peers exchanging their group status with each other after some time and queries being routed to RV peers. By arranging the RV peers into a group the communication between RV peers regarding service is localized and secure.

In Figure(4) a EPM group is shown having two RV peers as its members. The RV to RV communication is done in this group.

## 5. CONSCIENTIA SERVICES

In CONSCIENTIA Framework, there are three types of services.
- Entry Point Service.
- Monitoring Service.
- Worker Service.

### 5.1. Entry Point Service

This service entertains a client upon its service request to the service group. An instance of this service is running on all rendezvous peers in the service group. This redundancy of service removes hotspots from the network. It also makes the service robust by distributing the point of failure.

All queries are first received by the entry point service. A client peer sends a query to the RV peer with which it is registered. The entry point service running on the RV peer entertains this query by distributing the query in the worker subgroups.

The distribution strategy involves the optimal selection of a worker peer. This is a two step process. First, the Entry point service selects the best worker group and routes the query to it. Second, it then routes the query to the best worker node.

Entry point service maintains a cache of scheduled queries. In case, of worker peer failure, the monitoring service informs the Entry Point Service about the failure. The Entry Point Service then reschedules the query to another worker according to query distribution strategy. This scenario raises the issue of entertaining a client twice, to solve this each worker when replying to a query sends the "query serviced" message to its RV peer. This removes the query from the schedule list. The updated list now holds only the unprocessed queries that are to be rescheduled.

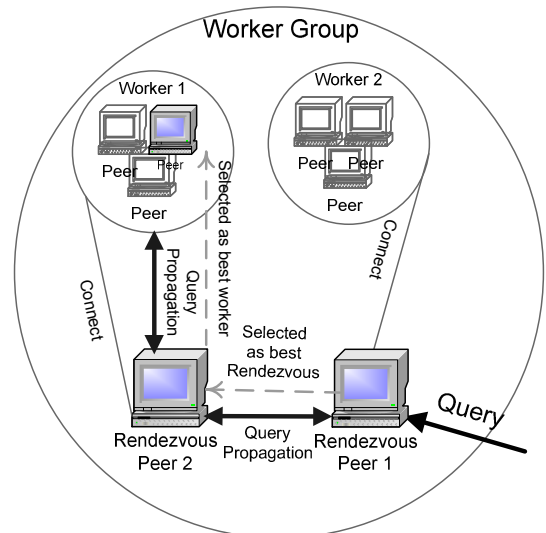

Figure (5) Query distribution

The failure of an entry point means the failure of a RV peer. Whenever a RV peer fails the worker peers in the subgroup compete to adopt the role of RV peer of that group. The first peer that becomes RV launches the Entry point service. The Entry point service in accordance with Monitoring service gets the lists of all the queries being processed in the network and rebuilds its scheduling record.

### 5.2. Monitoring Service

The monitoring service provides facilities for the self healing and self managing aspect of the framework.

#### 5.2.1. Monitor service in Worker Group

The Monitoring service has a module local to each subgroup, meaning that all the RV peers are running this service at their respective worker subgroup level.

All worker peers send a heart beat message to the Monitoring service running on their RV. The message encapsulates (1) timestamp and (2) the current thresh hold of the worker peer. Using this information and information from the Entry point service, the Monitoring service updates a table that demonstrates the (1)load, (2)queries scheduled and (3)the network delay of each peer.

#### 5.2.2. Monitoring service in EPM Group

All the RV peers exchange their tables after some time, thus at any given instance all the RV peer hold the complete picture of the service for the previous time slot. Thus in case of the failure of an entry point the effected group gets updated to the last good state. The worker peers hold their communication until connected to a rendezvous and so the worker sub group preserves it state until the entry point recovers.

In case of failure of a core service the Monitoring service makes new instances of that service to maintain the service. In case of resource exhaustion the Monitoring service create instances of the resource on available rendezvous and worker peers to maintain service quality and availability.

When a Monitor service fails then it means that a rendezvous peer has died. When the Entry point service of the effected group recovers, the monitoring service automatically recovers.

### 5.3. Worker Service

Worker service is hosted on worker peers. It provides the actual service an interface to CONSCIENTIA. It is an XML based interface which defines the query-format and reply-format. It is responsible for providing individual services i.e. receiving queries, processing them, generating response and sending them to the querying client.

A worker service automatically performs load balancing by dynamically adjusting a threshold value defined as

$$T_{new} = f(T_{old}, Q_{(x)}) - (T_{old} - Q_{(x)}) \quad ; \text{ if } (Q_{(x)} < T_{old})$$
$$\quad f(T_{old}, Q_{(x)}) - floor(T_{old}/10); \text{ if } (Q_{(x)} = T_{old}) \quad (1)$$

*where,*

$f(T_{old}, Q_{(x)}) = |T_{old} - 2Q_{(x)}|$

$Q_{(X)}$, *is the number of queries processed in the previous X seconds.*

Thus worker service dynamically adjusts to the network and service load.

Whenever a worker service receives a query it extracts the client peer's Pipe ID and the query. The query is sent to the actual service which is hosted on the worker peer. The response receive is sent to the client peer via its Pipe.

The worker service sends its threshold and time stamp to the monitoring service running on its RV peer. Whenever a query is processed a "query processed" message is sent to the Entry point service to update the schedule table.

If the connection with a rendezvous is lost then the worker waits for some time, if no rendezvous is found then it makes itself the rendezvous and advertises its new status. During this wait all the query acknowledgement messages and replies are cached, this preserves the state of the peer and in turn the whole group.

### 6. APPLICATIONS

To a service application the CONSCIENTIA framework appears as three layered stack.

The bottom layer is the CONSCIENTIA Core layer. It runs the core services that provide self healing, self managing mechanism to the peers, and an interface between the client and the service.

The middle layer is the peer layer; each peer adopts a specific role according to the services running on it.

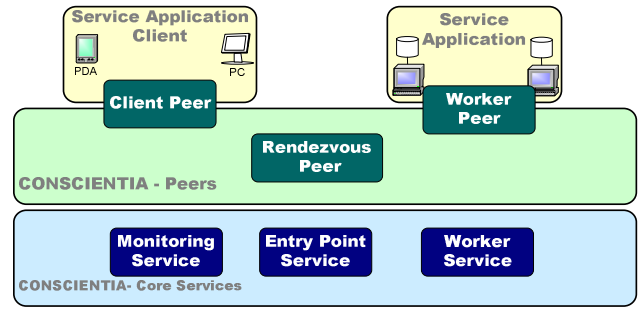

Figure (6) CONSCIENTIA Layered Architecture

The top layer is the client and service application layer. The clients and service applications connect to peers and use the service running on them to utilize the framework.

This framework is suitable for applications that demand a distributed, decentralized and autonomous environment. CONSCIENTIA can be used provide the following services:
- Distributed data management for Scientific and commercial applications.
- Collaborative Work (Web services, Ad Hoc Networking, Multiplayer Games, etc).
- Share CPU and other Resources.
- Data warehousing services independent of the underlying database.
- Implementing Distributed metadata catalogs.

### 7. CONCLUSION

Conscientia is a multi stage effort to provide self healing, self managing and scalable framework for collaborative hybrid services. Conscientia achieves this by using the redundant and loosely coupled nature of P2P networks, Compliance with service oriented architecture (SOA), Platform independence of XML and Java.

It is a platform independent, network independent and language independent framework, capable of extending a service to every device with a digital heartbeat. In Conscientia a service needs to be made only once, and then clients can use any possible mechanism to use it, the JXTA TINI Project[6] will enable any electrical device to become a peer and contribute in Conscientia.

### 8. ACKNOWLEDGMENTS

We will like to thank Foundation University Islamabad Pakistan, National University of Sciences and Technology, Pakistan and Center for European Nuclear Research Switzerland, for their generous technical and moral support, motivation and encouragement for this project.